\documentclass[prb,twocolumn,amscd,amsmath,amssymb,letterpaper]{revtex4-1}

\usepackage{graphicx} 
\usepackage{epstopdf} 
\usepackage{cancel} 
\usepackage{color}
\usepackage{amsthm}

\begin{document}
\title{Cooling of cryogenic electron bilayers via the Coulomb interaction }
\author{John King Gamble}
\email{jgamble@wisc.edu}
\author{Mark Friesen, Robert Joynt}
\author{S. N. Coppersmith}
\email{snc@physics.wisc.edu}
\affiliation{University of Wisconsin-Madison, Physics Department \\ 1150 University Ave, Madison, Wisconsin 53706, USA}

\date{\today}

\begin{abstract} 
Heat dissipation in current-carrying cryogenic nanostructures is problematic because the phonon density of states decreases strongly as energy decreases. We show that the Coulomb interaction can prove a valuable resource for carrier cooling via coupling
to a nearby, cold electron reservoir. Specifically, we consider the geometry of an electron bilayer in a silicon-based heterostructure, 
and analyze the power transfer. We show that across a range of temperatures, separations, and sheet densities, the
electron-electron interaction dominates the phonon heat-dissipation modes as the main cooling mechanism. Coulomb cooling is most effective at low densities, when phonon cooling is least effective in silicon, making it especially relevant for experiments attempting to perform coherent manipulations of single spins.

\end{abstract} 

\maketitle

\section{Introduction}
As researchers continue to probe smaller electronic devices at lower temperatures, a detailed understanding of heat management 
applicable on such length and energy scales becomes increasingly important. 
For example, recent experiments to detect the spin resonance of a single electron \cite{xiao} and 
to perform fast charge sensing in few-electron quantum dots \cite{PhysRevB.81.161308} are both limited by
heating effects.
Other applications, such as the search for the $\nu = 5/2$ non-abelian quantum Hall state, are expected to require
very low temperatures,\cite{2008arXiv0802.0930D} making the development of schemes for cooling such devices a necessary challenge.

The main problem is that whenever current is applied to a device to perform a measurement, the conduction electron temperature increases
due to Joule heating.\cite{Simmons:2010p1466} 
In devices operating near room temperature, heat can be readily dissipated through phonons, as the conduction electrons and lattice are strongly coupled.
However, as temperature is decreased, the conduction electrons decouple from the lattice. 
The phonon modes contribute less and less to cooling because the phonon density of states decreases as energy
is decreased.\cite{ashcroft_mermin}
Hence, as the system gets colder, it becomes more difficult to cool via conventional means.

Besides phonon cooling, systems can be cooled by electron diffusion through the leads.\cite{Prus:2002p90}
However, in common nanoscale devices, the leads extend hundreds of microns from critical regions to regions that are well cooled. 
This large distance scale limits the effectiveness of electron diffusion for device cooling.\cite{Prus:2002p90}

Here, we investigate using the Coulomb interaction directly, a strategy for cooling that remains largely unexplored. That
is,  we consider placing a cold conductor nearby the hot conduction electrons, and using  the Coulomb interaction for heat transfer. The
cold body could then be heat-sunk directly, without having to worry about interfering with the operation of a device.

While it may seem that remote Coulomb interactions are not strong enough to facilitate meaningful power transfer, 
several recent experiments have shown that remote interactions can indeed drastically affect electron relaxation. For instance,
the widely studied Coulomb drag (CD) effect \cite{PhysRevLett.88.126804,Jauho:1993p95,PhysRevB.48.8203} involves the transfer of momentum 
from one two-dimensional electron gas (2DEG) to another via the Coulomb interaction, due to the layers' close proximity.
Another example of the importance of remote Coulomb interactions arises in the metal-oxide-semiconductor (MOS) geometry, where it
has been found that device performance can be reduced due to interactions of the conduction electrons
with those in the gate when the distances are too small.\cite{Fischetti:2001p1358,Fischetti:2001p1359,Laikhtman:2005p1355}

In this paper, we consider two parallel, silicon 2DEGs, separated by tens of nanometers.
We make the simplifying approximation that  both 2DEGs are of zero thickness.
Similar devices have been implemented experimentally in the form of silicon electron-hole bilayers.\cite{Prunnila:2008p1269,Takashina:2009p1354}
One of the layers is taken to have a temperature on the order of tens to hundreds of mK; this is the active layer that
we are interested in cooling.
The second layer is taken to be a heat sink, and is assumed to be at temperature $T = 0$ K.
Although cooling this heat sink layer would be a challenging engineering problem, electron diffusion through grounded, close proximity
leads can effectively cool samples to about $10$ mK.\cite{PhysRevLett.103.236802} 
Since the two layers are electrically decoupled, this would not interfere with electrical measurements on the active layer. 

We study the temperature, separation, and carrier density dependencies of the heat transfer, and compare
it to experimental results for heat dissipation due to phonons. 
The Coulomb interaction is found to be competitive and even dominant over phonons for a range of
temperatures and densities, for separations up to several tens of nanometers.
Specifically, we find that lowering the electron density enhances the power transfer due to the Coulomb interaction, but decreases the power dissipation due to phonons.

This paper is organized as follows. 
Section~\ref{sec:formalism} uses a formalism similar to that used for CD to formulate the physical problem.
Next, Sec.~\ref{results} finds an expression for the power transfer and discusses its asymptotic behavior in both near and far distance regimes.
Section \ref{phonons} compares the results of the previous sections  to experimental results for phonon-mediated cooling.
Finally, Sec.~\ref{sec:discussion} discusses the implications of cooling nanosystems using carrier-carrier interactions, and suggests
directions for future study.

\section{Formulation of the problem} \label{sec:formalism}
In this section, we describe the physical situation we will consider throughout this paper: two parallel 2DEGs, one serving as a heat sink for the other.
We then review the standard scattering formalism that is used to perform Coulomb drag (CD) calculations. Using this Boltzmann transport formalism,
we write down an equation for power transfer, which we will evaluate in subsequent sections.

The physical situation we consider here is very similar to that which has been well-studied in the CD literature.\cite{PhysRevLett.88.126804,Jauho:1993p95,PhysRevB.48.8203,Rojo:1999p10} Specifically, we have a sample that contains two 2DEGs that are parallel but spatially separated by some distance $d$. In the case of CD, one of the 2DEGs is driven
by a current, while the other is not. 
Because of the current flow, the distribution function in the first layer is out of equilibrium, and the resulting charge fluctuations generate a response in the second layer, mediated
by the Coulomb interaction between the layers. Although complicated by screening, the basic picture is that the presence of the current in the first layer ``drags" electrons in the
second layer, creating a net voltage.

In our case, we do not consider the non-equilibrium effects of a current flowing. Rather, we suppose that each layer is internally at thermal equilibrium, but at different temperatures.
We will consider one layer (the active layer) to be at a finite temperature $T$, and the second layer (the heat sink) to be at $T=0$ K.
Intuitively, we expect that energy should be transferred from the active layer to the heat sink. Microscopically, this is due to density fluctuations in the hot layer causing responses in the 
cold layer, mediated by the Coulomb interaction. Fortunately, we can borrow much of the initial setup of the problem from the CD formalism. However, the evaluation of the 
resulting expression is quite different because we focus on energy between 2DEGS of unequal temperature, rather than momentum transfer due to a driving electric field.

Since we will be working at very low temperatures ($<1$ K),  binary Coulomb collisions are the primary method of heat transfer between layers. If we were to consider temperatures $T \gtrsim 0.2 E_F/k_B$, where $E_F$ is the Fermi energy, we would also need to take into account collective scattering effects, the so-called plasmon enhancement.\cite{Rojo:1999p10}
Hence, we consider interactions that transfer energy from an electron in the active 2DEG (layer 1) to an electron in the heat sink 2DEG (layer 2).
The electrons involved in the interaction have initial (2D) momenta $\left( \mathbf k_1,\mathbf k_2 \right)$ and final momenta $\left( \mathbf k_1',
\mathbf k_2'\right) = \left( \mathbf k_1 + \mathbf k,
\mathbf k_2 - \mathbf k\right)$, where $\mathbf k$ is the transferred momentum. 
This carrier-carrier scattering falls into the category of distinguishable particle scattering since events only occur between
particles in different layers.
The formalism for treating such a scattering problem is well known,\cite{ridley} and the power transfer is shown in appendix \ref{energy_transfer} to be
\begin{equation}\label{eq:formal_energy_transfer}
P =  \frac{16 A^3}{(2 \pi )^6} \int d^2k_1 d^2 k_2 d^2k \cdot E  \cdot \Gamma,
\end{equation}
where $A$ is the sample area, $E$ is the transfer energy for an individual event, and $\Gamma$ is the scattering rate. 
In the next section, we evaluate Eq.~(\ref{eq:formal_energy_transfer}) using the Thomas-Fermi approximation to describe screening (which in this context is equivalent to the RPA approach taken in Ref.~\onlinecite{Rojo:1999p10}).

 \begin{figure}[tb!] 
\begin{center} 
\includegraphics[width=1.0 \linewidth]{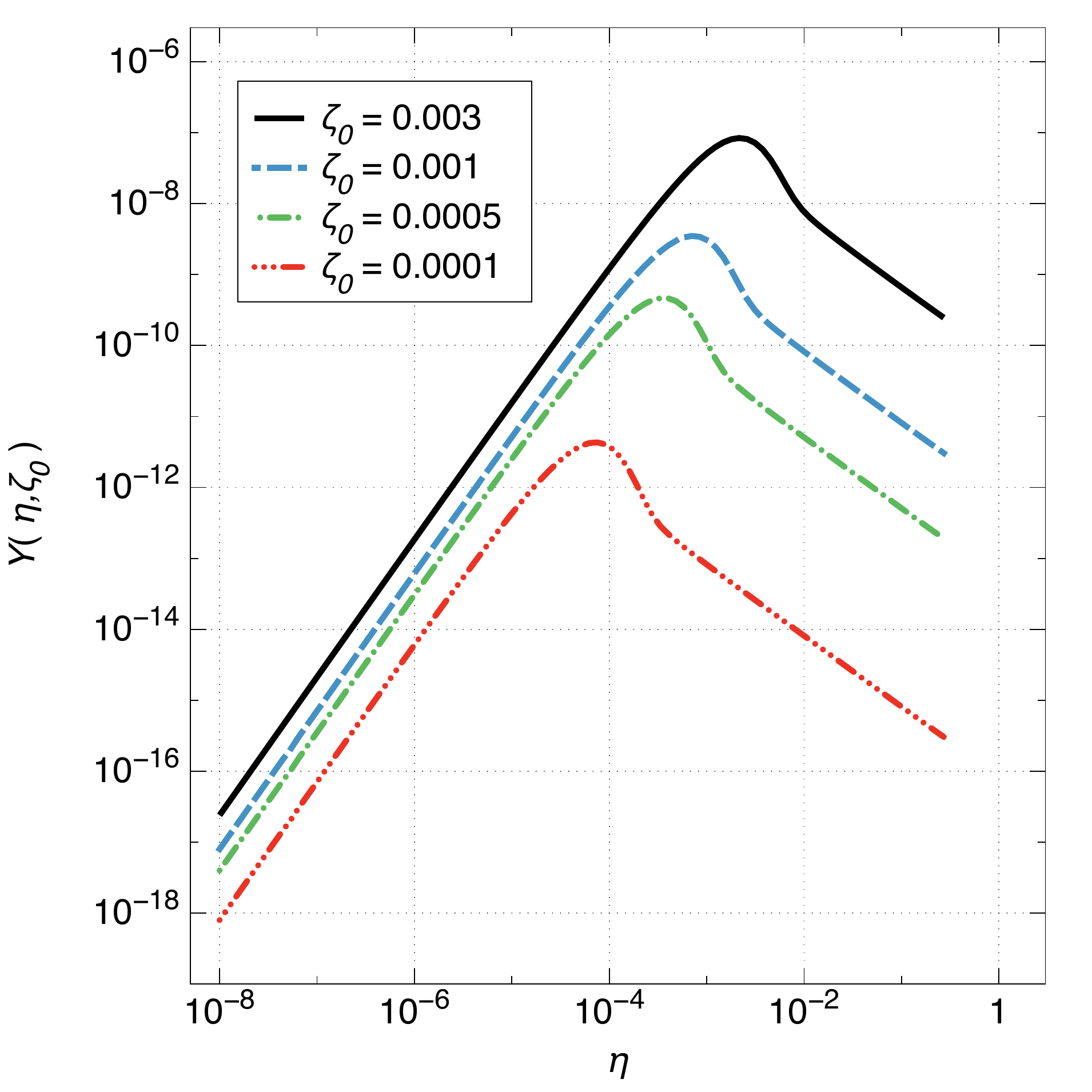}
\end{center} 
\caption{(Color online) The dimensionless function $Y(\eta,\zeta_0)$ (Eq.~(\ref{yfunc2})) plotted over $\eta$ for various values of $\zeta_0$ and with layer densities $n_1 = n_2$. Here, $\eta = k/k_F$ is a scaled momentum and $\zeta_0 = k_B T / E_F$ is the characteristic magnitude of energy fluctuations due to temperature. The function follows a power law for both low and high $\eta$, with a maximum occurring
for an intermediate $\eta$, which we denote as $\eta^*$.
\label{fig_1} }
\end{figure}

\section{Results}\label{results}
In this section, we evaluate Eq.~(\ref{eq:formal_energy_transfer}), using some standard methods incorporated into the 
calculation of CD. However, the resulting expression is quite different, so we work through its derivation in appendix~\ref{energy_calculation}.
We set the temperature of the active layer to $T$, the temperature of the heat sink layer to absolute zero, the Fermi level of the active layer to $E_F$, and 
the Fermi level of the heat sink layer to $E_F/x$, where $x=n_1/n_2$, the ratio of carrier densities between layer one and layer two. The power transfer is then
\begin{equation}\label{PbyAversion1}
\frac{P}{A} = \frac{E_F^4}{64 \hbar} \left( \frac{\epsilon_0 \epsilon_b}{q^2} \right)^2 \int d \eta  \left( \frac{\eta}{\sinh(\eta / \eta_0)} \right)^2 Y(\eta, \zeta_0),
\end{equation}
where 
\begin{align} \label{yfunc2}
Y(\eta,\zeta_0) &= \int_0^{\infty} d \zeta  \frac{\zeta}{\eta^3}  \left[  \coth \left( \zeta/\zeta_0 \right)-1 \right] \nonumber \\
& \times \mathrm{Re} \Big(\sqrt{ 2(2 + \zeta) \eta^2 - \eta^4 - \zeta^2} \nonumber \\
& -\sqrt{ 2(2 - \zeta) \eta^2 - \eta^4 - \zeta^2}  \Big) \nonumber \\
& \times \mathrm{Re} \Big(\sqrt{ 2\left(\frac{2}{x} + \zeta\right) \eta^2 - \eta^4 - \zeta^2} \nonumber \\
& -\sqrt{ 2\left(\frac{2}{x} - \zeta\right) \eta^2 - \eta^4 - \zeta^2}  \Big),
\end{align} 
and the dimensionless parameters are $\zeta \equiv E/E_F$, $\eta \equiv k/k_F$,  $\zeta_0 \equiv k_B T/E_F$ and $\eta_0 \equiv 1/(k_F d)$,
with $k_F = \sqrt{2 m^* E_F}/\hbar$ the Fermi momentum. Here, the $\eta/\sinh(\eta/\eta_0)$ term is due to the interlayer Coulomb interaction (Eq.~(\ref{screened_2d_pot})), while 
the distribution functions give rise to $Y(\eta,\zeta_0)$.

The function $Y(\eta,\zeta_0) $ is plotted in Fig.~\ref{fig_1}, where it
is shown that $Y$ is a peaked function in $\eta$ varying as a power of $\eta$ on either side of the peak.
We define $\eta^*$ to be the location of the peak, and note that $\eta^* \approx \zeta_0$.
Physically, $Y(\eta,\zeta_0)$ tracks the availability of energy fluctuations corresponding to a particular momentum transfer
$k = \eta k_F$ and temperature $ k_B T = \zeta_0 E_F$. 
If $\eta<\eta^*$, $Y$ is limited by the Fermi-Dirac distributions that govern the occupation of states in each 2DEG.
For $\eta>\eta^*$, $Y$ is instead constrained by the temperature difference between 2DEGs.

From Eq.~(\ref{PbyAversion1}), we see that the Coulomb potential causes $\eta$ to be cut off at approximately $\eta_0$. 
Hence, there are two asymptotic regions of interest: when $\eta_0\ll\eta^*$ and when $\eta_0\gg\eta^*$.
In the first region, the Coulomb potential cuts off the integration over $\eta$ well before $\eta^*$,
which corresponds to large separations between 2DEGs.
Here, the separation distance limits the magnitude of the momentum transfer, which in turn limits power transfer.
For $\eta_0 \gg \eta^*$, the Coulomb interaction truncates the $\eta$ integration after $\eta^*$, corresponding to small separation.
In this regime, $Y$ is already rapidly decreasing, so the power transfer is instead mainly constrained by the temperature difference between 2DEGs.
The crossover between these two regions occurs when $\eta_0 \approx \eta^*\approx \zeta_0$, corresponding
to a separation of $d \approx E_F/(k_F k_B T)$.

 \begin{figure}[tb!] 
\begin{center} 
\includegraphics[width=1.0 \linewidth]{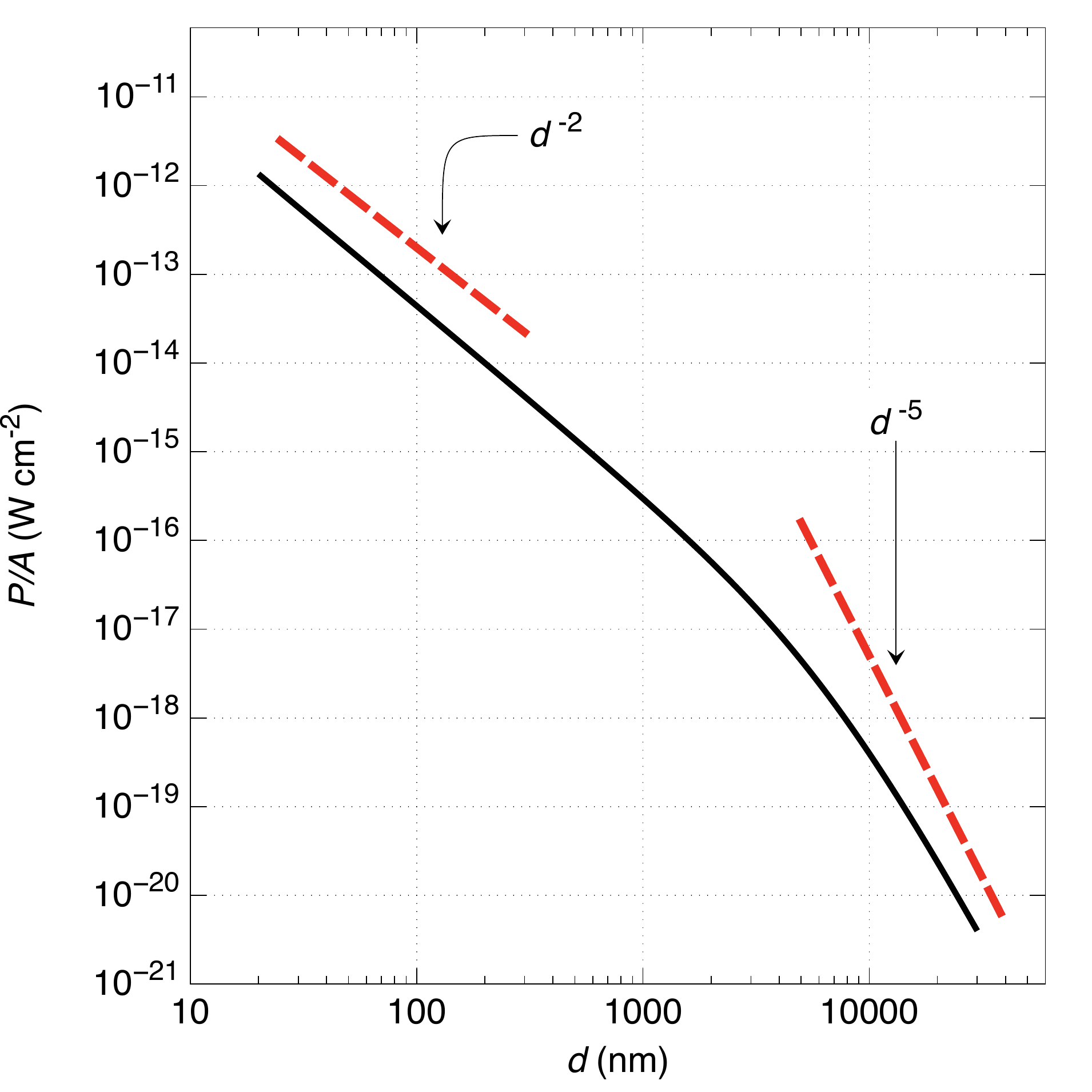}
\end{center} 
\caption{(Color online) Calculated values for the power per unit area $P/A$ transferred via the Coulomb interaction versus separation  (Eq.~(\ref{PbyAversion1})). Here, the 
sheet density is $4 \times 10^{11}$ cm$^{-2}$ in both layers, the temperature of the active layer is $T=100$ mK, and the temperature of the heat sink layer is 0 K. As can be seen by the dashed lines, the power transfer varies as approximately $1/d^2$ at small distances and  $1/d^5$ at large distances. The crossover length scale occurs when $d \approx  E_F/(k_F k_B T)$.
\label{fig_2} }
\end{figure}

In appendix~\ref{asymptotic}, we work out the asymptotic forms for power transfer. We consider the specific case of equal density 2DEGs, when $x=1$. In the large separation regime when $d \gg  E_F/(k_F k_B T)$, we find
\begin{equation}\label{asymptotic_large_d}
\frac{P}{A} \sim \frac{k_F \hbar^5}{512 {m^*}^3}  \left( \frac{\epsilon_0 \epsilon_b}{q^2} \right)^2 \frac{k_B T}{d^5}
\left[ 8.3 + 13.0 \cdot \log\left( k_F d \right) \right],
\end{equation}
where we use $(\sim)$ to denote asymptotic equivalence.
For the short distance limit where $d \ll  E_F/(k_F k_B T)$, we have
\begin{align}\label{asymptotic_small_d}
\frac{P}{A} &\sim \frac{\hbar}{128 E_F m^*}  \left( \frac{\epsilon_0 \epsilon_b}{q^2} \right)^2 \frac{k_B^4 T^4}{d^2} \nonumber \\
& \times \left[ 0.46-1.32\cdot \log\left( \frac{k_B T}{E_F}\right) - 0.81 \cdot \log\left(k_F d \right)\right].
\end{align}
Hence, up to logarithmic corrections, $P/A \propto T/d^5$ for large distances and $P/A \propto T^4/d^2$ for small distances, which can be 
qualitatively understood as follows. At low temperatures ($T \ll k_B E_F$), it is reasonable to assume that energy
fluctuations are small and concentrated about the Fermi level, so that the transfer momentum obeys $k \ll k_F$. Expanding Eq.~(\ref{yfunc2}) for small $\eta$, while working in the large separation regime where $\eta\ll\eta^*$, we find that
\begin{equation}
Y(\eta,\zeta_0)  \propto \zeta_0 \eta^2.
\end{equation}
Here, the scaling is determined by the Fermi-Dirac distributions limiting the power transfer.
Likewise, if we work in the small separation region where $\eta\gg\eta^*$, we find
\begin{equation}
Y(\eta,\zeta_0)  \propto \frac{\zeta_0^4}{\eta},
\end{equation}
where the scaling is now determined by the layer temperature.
In these limits, since the Coulomb interaction sets the scale of $\eta \propto \eta_0$, we can easily see the rough dependences (neglecting the logarithmic corrections) via power counting in Eq.~(\ref{PbyAversion1}).
Fig. \ref{fig_2} shows the numerical evaluation of Eq.~(\ref{PbyAversion1}) as a function of separation, clearly demonstrating both distance regimes. 
 
\section{Comparison with cooling due to phonons.}\label{phonons}
In this section, we compare Coulomb-mediated cooling to experimentally measured energy dissipation due to phonons at low temperatures. 
In silicon-based heterostructures at low temperatures, two types of phonon couplings are important: acoustic phonons governed by a deformation potential coupling,\cite{Karpus:1990p1462} and the Pekar coupling.\cite{PhysRevB.71.081305}
Pekar phonons arise from the sharp electrostatic confinement potentials present in heterostructure devices, such as quantum wells, and hence
are not present in bulk samples.\cite{PhysRevB.71.081305}
They also share a characteristic $T^3$ dependence with piezoelectric phonons,\cite{PhysRevB.71.081305} making them especially important in low-temperature experiments with non-polar materials, such as few-electron quantum dots in Si.

Indeed, both deformation potential and Pekar phonons have been experimentally observed in silicon-based heterostructures at low temperature.\cite{Prus:2002p90}
The characteristic temperature dependence for deformation potential coupling is $T^5$,\cite{Karpus:1990p1462}
so for very low temperatures we expect Pekar phonons to dominate, while for higher temperatures deformation potential phonons become more important. 

As established in Eq.~(\ref{asymptotic_small_d}), for small separations the power transfer to the heat sink layer via the Coulomb interaction varies as $T^4$. Whether or not this Coulomb cooling is larger than phonon cooling over a given temperature range depends on the numerical magnitude of Eq.~(\ref{PbyAversion1}), which we now compute. 
We compare Coulomb cooling to experimental measurements of phonon mediated cooling in Ref.~\onlinecite{Prus:2002p90}. 
There, it is found that the power dissipation due to phonons is $P_{ph}/A = aT^3 + b T^5$, where $a = 2.2 \times 10^{-8}$ W~K$^{-3}$~cm$^{-2}$
and $b = 5.1 \times 10^{-8}$ W~K$^{-5}$~cm$^{-2}$. The structure used is a silicon MOS inversion layer, with dielectric thickness $200$ nm and
carrier density $5.4 \times 10^{11}$ cm$^{-2}$. 

It is known that the phonon 
couplings depend on the electron density, with $P \propto n^{-3/2}$ for the deformation potential coupling.\cite{Karpus:1990p1462} Pekar phonons have both an explicit $n^{-1/2}$ dependence and a dependence on the electric field at the 2DEG of $F^2$.\cite{PhysRevB.71.081305} Since in a 2DEG $F \propto n$,\cite{davies} Pekar phonons scale as $P \propto n^{3/2}$ in total.
By comparison, Eq.~(\ref{asymptotic_small_d}) tells us that for equal density 2DEGs, Coulomb power transfer goes like $P \propto 1/n$. 

\begin{figure}[tb!] 
\begin{center} 
\includegraphics[width=1.0 \linewidth]{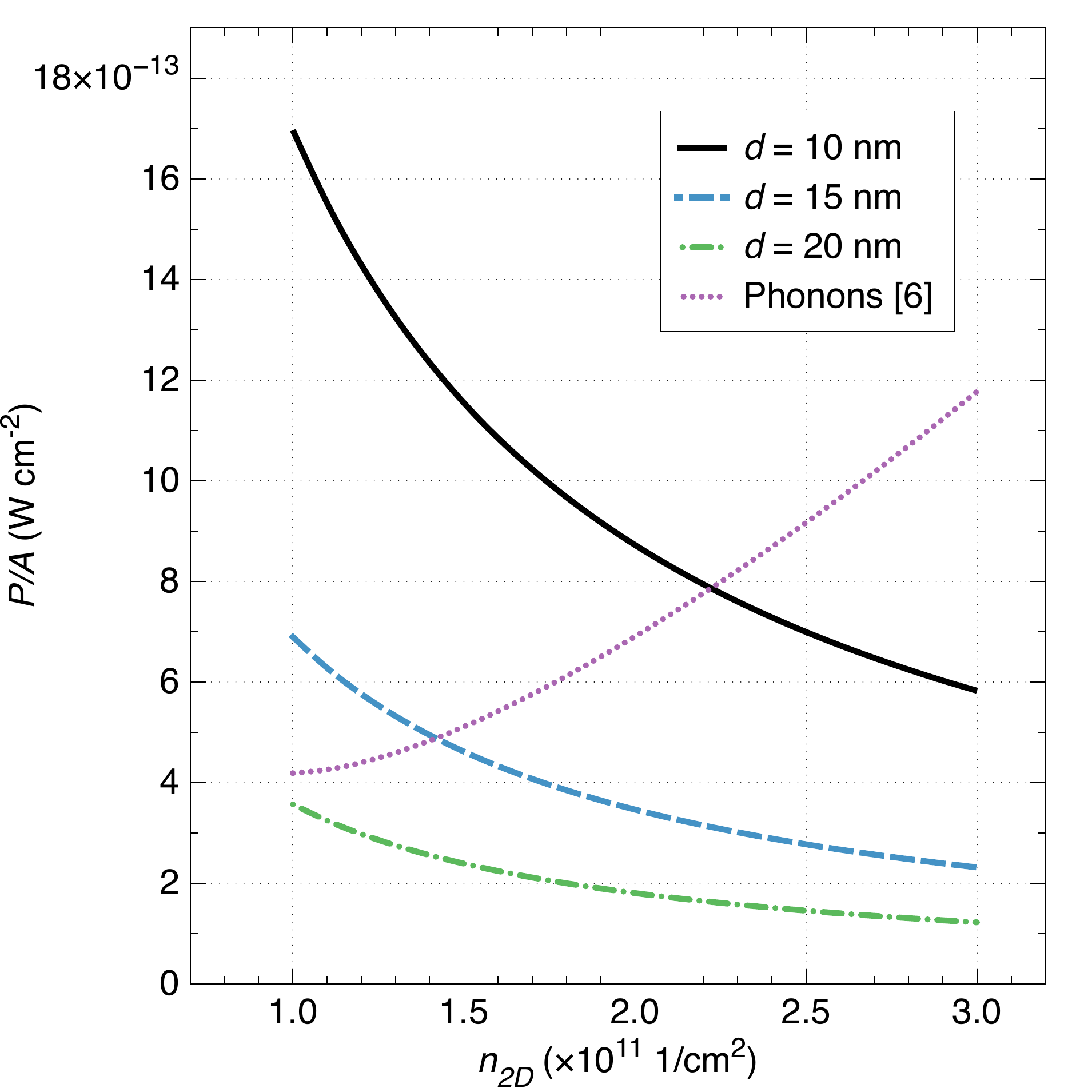}
\end{center} 
\caption{(Color online) Calculated values for the power per unit area $P/A$ transferred 
between an active ($T=50$ mK) and heat sink ($T=0$ K) 2DEG
via the Coulomb interaction as a function of the layer density at three different values for the separation between layers. Here, the densities of both layers are identical. 
For comparison, $P/A$ due to phonons from experimental data in Ref.~\onlinecite{Prus:2002p90}, scaled for changing density, is shown as a dotted line. \label{fig_3}}
\end{figure}

Typically, in an experiment the density is fixed by desired electronic properties (for instance, the ability to pinch off current
with depletion gates). 
For low-temperature applications that attempt to reach few-electron regimes, it  is desirable to have a low density.
It is therefore important to determine the dependence of power transfer on the density $n$.
Fig.~\ref{fig_3} shows the effect of varying the layer density on the power transfer, for three different layer separations at constant temperature $T=50$ mK. As expected, the Coulomb power transfer is greatest in the case of small density, making it especially pertinent for few-electron experiments. 

 It is important to note that our formalism for static screening is only valid when the transfer momentum obeys $k < 2 k_F$,\cite{davies} which means that our we cannot make the density too small. The Coulomb interaction limits the transfer momentum to $k \lesssim 1/d$. Hence, setting $k = 1/d$ for $d = 10$ nm corresponds to $n > 0.79 \times 10^{11}$ cm$^{-2}$. Another constraint on low-density 2DEGs is the 
 metal-insulator transition, which occurs for sufficiently low densities. In silicon MOS structures, 
 the critical value of density is known to be around $n_c \approx 1 \times 10^{11}$ cm$^{-2}$.\cite{dasSarma}
 More recently, calculations for dopantless Si/SiGe devices predict that this value can be much lower, about  $n_c \approx 2 \times 10^{10}$ cm$^{-2}$.\cite{gold}

\begin{figure}[tb!] 
\begin{center} 
\includegraphics[width=1.0 \linewidth]{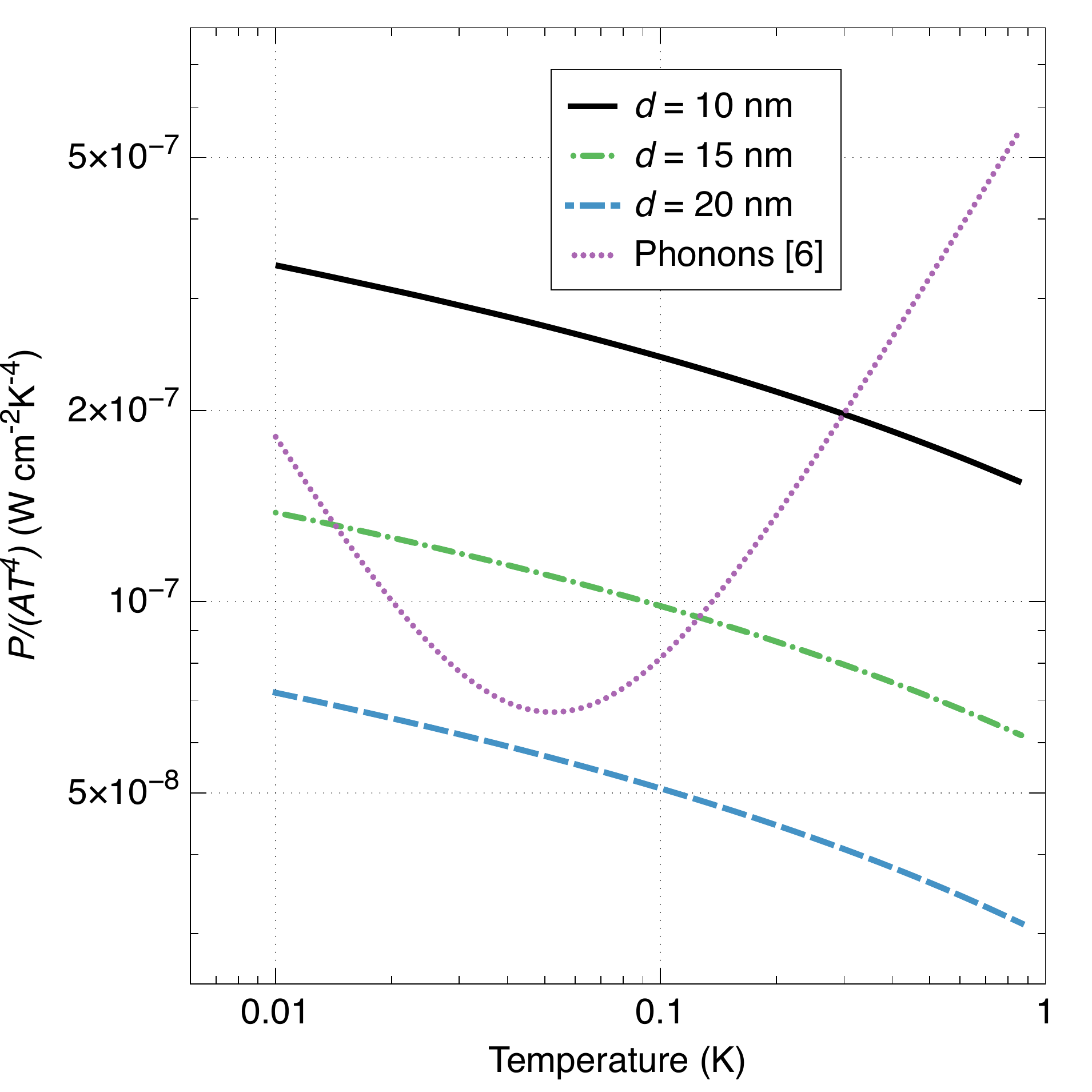}
\end{center} 
\caption{(Color online) Scaled values for the power per unit area $P/(AT^4)$ transferred via the Coulomb interaction 
between two 2DEGs versus temperature at three different values for the separation between the layers.
For comparison, the scaled $P/(AT^4)$ due to phonons from experimental data in Ref.~\onlinecite{Prus:2002p90} is shown. The density of
both layers is $1\times 10^{11}$/cm$^2$. \label{fig_4} }
\end{figure}

In Fig.~\ref{fig_4}, we plot the temperature dependence of the power transfer per unit area $P/A$ for several separations, and compare with the power dissipation due
to phonons. There, we fix the carrier density to be $n=1\times 10^{11}$ cm$^{-2}$ for both layers. One sees that for small separations (less than 20 nm), Coulomb-mediated power transfer exceeds phonon
power dissipation over a potentially wide temperature range (roughly up to 300 mK for a 10 nm separation).

\section{Discussion}\label{sec:discussion}
Understanding relevant heat dissipation mechanisms at low temperatures in electronic devices is an important problem, especially as
spin-based, few-electron devices mature. In this paper, we considered a geometry consisting of parallel 2DEGs in silicon and
calculated the expression for power transfer between two layers at temperatures $T>0$ and $T=0$ respectively, in the approximation of Thomas-Fermi screening. 
We then presented analytic results for the asymptotic regimes of small and large separations.
We showed that in this geometry, power transfer due to the remote Coulomb interaction can be the dominant heat loss mechanism. 
This Coulomb cooling is most effective at low densities, making it especially important for experiments attempting to access few-electron regimes.

There have been a number of studies of heat transfer between close bodies, including a semiclassical kinetic treatment 
by Boiko and Sirenko \cite{PSSB:PSSB2221590228} and an electromagnetic formulation by Volokitin and Persson.
\cite{Volokitin:2007p578,Volokitin:2001p609} However, these are largely interested in hot devices, where
complicating features such as plasma excitations are important. Further, as noted in Ref.~\onlinecite{Volokitin:2007p578}, 
there are discrepancies between this electromagnetic formalism and Boltzmann transport approaches.
More recent work by Kr\"uger, Emig, and Kardar extends the electromagnetic formalism to arbitrary geometries with a focus on heat transfer.\cite{kruger}
It would be beneficial to compare the present work to the electromagnetic treatments to attempt to address the origin of
any discrepancies.

While the results for two parallel 2DEGs are promising, one could almost certainly engineer a better geometry for
optimizing heat dissipation. Indeed, the main reason for a preliminary evaluation of the 2DEG-2DEG geometry was
due to its computational simplicity. An idea for a more effective heat sink might be a standard MOS geometry, or 
a top-gated nanostructure. Due to the drastically higher density of states in the metal, one could expect an enhanced 
power transfer. However, screening would also be enhanced, so careful calculations, similar to those presented
 in this paper, should be done for that geometry. Also, studying the effects of high-k dielectrics might be fruitful, since
 the power transfer scales as $\epsilon_b^2$.

\section{Acknowledgements}
The authors thank A. L. Saraiva and M. A. Eriksson for useful discussions.
This work was supported in part by ARO and LPS (W911NF-08-1-0482), NSF (DMR-0906951), and NSF (DMR-0805045).
JKG gratefully acknowledges support from the National Science Foundation.
\appendix

\appendix

\section{Derivation of the power transfer rate}\label{energy_transfer}
In this appendix, we briefly sketch the derivation of Eq. (\ref{eq:formal_energy_transfer}), the formal expression for the power transfer between two 2DEGs, using the methods of Ref.~\onlinecite{ridley}.
First, recall that we are interested in the scattering of two particles with initial (2D) momenta $\left( \mathbf k_1,\mathbf k_2 \right)$ and final momenta $\left( \mathbf k_1',
\mathbf k_2'\right) = \left( \mathbf k_1 + \mathbf k,
\mathbf k_2 - \mathbf k\right)$.
The transition rate $\Gamma$ for the above process is given by the
balance equation
\begin{align}\label{gammafunction}
\Gamma\left( \mathbf k_1,\mathbf k_2;\mathbf k_1',\mathbf k_2' \right) &= S \left( \mathbf k_1,\mathbf k_2;\mathbf k_1',\mathbf k_2' \right) \\
& \times   \Big[
f^{(1)}_{\mathbf k_1} \left( 1- f^{(1)}_{\mathbf k_1' }\right) f^{(2)}_{\mathbf k_2}  \left( 1- f^{(2)}_{\mathbf k_2' }\right) \nonumber \\
&-f^{(1)}_{\mathbf k_1'} \left( 1- f^{(1)}_{\mathbf k_1 }\right) f^{(2)}_{\mathbf k_2'}  \left( 1- f^{(2)}_{\mathbf k_2 }\right) 
\Big] , \nonumber 
\end{align}
where $S$ is the transition rate given that the appropriate states are available, $f^{(1)}$ is the Fermi-Dirac distribution function in layer one, 
\begin{equation}
f^{(1)}_{\mathbf k} = \left[ 1 + \exp \left( \frac{E_{\mathbf k} - E_F}{k_B T} \right) \right]^{-1},
\end{equation}
where $E_{\mathbf k} = \hbar^2 k^2/(2 m^*)$, $E_F$ is the Fermi level, $m^*$ is the effective mass, and $T$ is the temperature in layer one, and $f^{(2)}_{\mathbf k}$ is likewise the Fermi-Dirac distribution function in layer two. Note that here we restrict our attention to Fermi-Dirac distribution functions, but Eq.~(\ref{gammafunction}) remains valid even for non-equilibrium distribution functions.

The first term in the square brackets of Eq.~(\ref{gammafunction}) can be understood as the particles starting with momenta $\left( \mathbf k_1,\mathbf k_2 \right)$ and ending with momenta $\left( \mathbf k_1',\mathbf k_2' \right)$.
The second term corresponds to scattering from momenta $\left( \mathbf k_1',\mathbf k_2' \right)$ to momenta $\left( \mathbf k_1,\mathbf k_2 \right)$. To calculate the scattering rate $S$, we use Fermi's golden rule:
\begin{align}
S\left( \mathbf k_1,\mathbf k_2;\mathbf k_1',\mathbf k_2' \right) & = \frac{2 \pi}{\hbar} \left| H \right|^2 \\
 & \times \delta \left( E_{\mathbf k_1} +E_{\mathbf k_2}
 - E_{\mathbf k_1'}-E_{\mathbf k_2'} \right), \nonumber 
\end{align}
where $H$ is the interaction matrix element,
\begin{equation}
H =  \frac{q}{A} \tilde \phi_{scr}(k,d),
\end{equation}
and $\tilde \phi_{scr}$ is the Fourier transformed screened Coulomb interaction between layers. 
Defining the Thomas-Fermi screening wavevector $k_{TF} = 2 m^*/(\pi \hbar^2) \cdot q^2/(2 \epsilon_0 \epsilon_b)$,
the Fourier-transformed screened Coulomb interaction between the layers, calculated within the Thomas-Fermi approximation is \cite{Rojo:1999p10}:
\begin{align} \label{screened_2d_pot}
\tilde \phi_{scr}(k,d) = \frac{k q }{4 k_{TF}^2 \epsilon_0 \epsilon_b} \frac{1}{\sinh(k d)},
\end{align}
where we have assumed $k \ll k_{TF}$.
For clarity of presentation, we present a self-contained derivation of this expression in appendix~\ref{coulomb_derivation}.

 Now that we have an expression for the scattering rate between particular states, we obtain the power transfer between the layers:
\begin{equation}
P =  16\sum_{\mathbf k_1, \mathbf k_2, \mathbf k} E  \cdot \Gamma\left( \mathbf k_1,\mathbf k_2;\mathbf k_1+ \mathbf k,\mathbf k_2 - \mathbf k \right),
\end{equation}
where the factor of $16$ is due to spin degeneracies of two and valley degeneracies of two in each electron layer, \cite{Schaffler:1997p178} and $E =  E_{\mathbf k_1+ \mathbf k} - E_{\mathbf k_1} $, the transferred energy. Converting the sum to an integral gives us Eq.~(\ref{eq:formal_energy_transfer}).

\section{The screened Coulomb potential}\label{coulomb_derivation}
In this appendix, we present a self-contained derivation of the screened interlayer Coulomb potential within the Thomas-Fermi approximation. 
Although this result can be obtained as a special case of the random phase approximation
result as described in Ref.~\onlinecite{Rojo:1999p10}, assuming a static screening formalism from the beginning
results in a considerably more transparent calculation. 
The technique we present here can also easily be implemented numerically to treat more complex geometries.

To start, we consider placing an electron into one of the 2DEGs. This results in an external, unscreened potential
$\phi_{ext}(\mathbf r,z)$ due to the external electron, where $\mathbf r$ is the 2D position within the plane of the 2DEG. The electron gas in both layers
can rearrange to screen this external charge, resulting in an induced potential $\phi_{ind}(\mathbf r,z)$. The screened potential
that an electron in the other layer feels is then $\phi_{scr}(\mathbf r,d) = \phi_{ext}(\mathbf r,d) +\phi_{ind}(\mathbf r,d)$, where we have
assumed that the two 2DEGs are separated by a distance $d$. Our objective is to calculate $\phi_{ind}$, from which we can compute $\phi_{scr}$.

We assume that our system is translationally invariant in the plane parallel to the 2DEGs, which we define to be the $x-y$ plane. It is convenient to exploit this translational invariance by taking a Fourier transform of the Poisson equation in the $x-y$ plane, yielding
\begin{equation} \label{transformed_PE_ind}
\left(\partial_z \epsilon(z) \partial_z - \epsilon(z) k^2 \right) \tilde \phi_{ind}(\mathbf k,z)  = - \tilde \rho_{ind}(\mathbf k,z),
\end{equation}
where we denote the Fourier transform of a function $f(\mathbf r,z)$ as
\begin{equation}
\tilde f(\mathbf k,z) = \int d^2 r f\left(\mathbf r,z\right) e^{- i \mathbf r  \cdot \mathbf k}.
\end{equation}
In Eq.~(\ref{transformed_PE_ind}), $\rho_{ind}$ is the induced charge density, responsible for the production of $\phi_{ind}$, and $\epsilon(z)$ is the dielectric function.
In a homogeneous medium, Eq.~(\ref{transformed_PE_ind}) has the general solution \cite{jackson}
\begin{equation}\label{phi_ind_1}
\tilde \phi_{ind}(k,z) = \frac{1}{2 k \epsilon_0 \epsilon_b} \int dz' e^{-k | z-z' |} \tilde \rho_{ind}(k,z'),
\end{equation}
so to find $\phi_{ind}$, we must calculate $\rho_{ind}$.

To determine $\rho_{ind}$, we first note that the total charge density $\rho_{tot}$ obeys
\begin{equation}
\rho_{tot}(\mathbf r, z) = \rho_0 + \rho_{ind},
\end{equation}
where $\rho_0$ is the charge density without an external charge present, and we have neglected the small density contribution from the external charge itself.
The dispersion relation for the electrons is given approximately by
\begin{equation}
E(\mathbf k,\mathbf r,z) \approx \frac{\hbar^2 k^2}{2 m^*} - q \phi_{scr}(\mathbf r,z),
\end{equation}
where $-q$ is the charge on an electron. By using the functional form of the Fermi-Dirac distribution, we can view charge density as a functional of Fermi energy:\cite{ashcroft_mermin}
\begin{equation}\label{eq:rhoindapprox}
\rho_{ind}(\mathbf r,z) \approx \rho_0\left(E_F+q \phi_{scr}\right) - \rho_0\left(E_F\right).
\end{equation}
Now, assuming that $q \phi_{scr} \ll E_F$, to first order in $\phi_{scr}$ Eq.~(\ref{eq:rhoindapprox}) is
\begin{equation}\label{TF_approx}
\rho_{ind}(\mathbf r,z) \approx -q^2 \frac{dn_0}{dE}\Big \vert_{E_F} \phi_{scr},
\end{equation}
where $\rho_0 = -q n_0$.
For low temperatures, $dn_0/dE\big \vert_{E_F} \approx g(\mathbf r, E_F)$, the local density of states evaluated at the Fermi level, which might vary spatially. 
For our geometry with two 2DEGs separated by a distance $d$, $g$ only varies in the $z$ direction: 
\begin{equation}
g(z) = g_{2D}  \left(\delta(z) + \delta(z-d) \right),
\end{equation}
where $g_{2D}$ is the energy-independent two-dimensional density of states.

Substituting Eq.~(\ref{TF_approx}) into Eq.~(\ref{phi_ind_1}), we find
\begin{align}\label{phi_ind_sol}
\tilde \phi_{ind}(z)&= -\frac{q^2 g_{2D} }{2 k \epsilon_0 \epsilon_b}\Big[ e^{-k |z|} \left(\tilde \phi_{ind}(0)+\tilde \phi_{ext}(0) \right) \nonumber \\
&+ e^{-k |z-d|} \left(\tilde \phi_{ind}(d)+\tilde \phi_{ext}(d) \right) \Big].
\end{align}
The external potential due to the external electron in the first layer satisfies \cite{davies}
\begin{equation}\label{phi_ext_simple}
\tilde \phi_{ext}(k,z) = \frac{q}{2 k \epsilon_0 \epsilon_b}e^{-k|z|}.
\end{equation}
Hence, evaluating Eq.~(\ref{phi_ind_sol}) for $z=0$ and $z=d$ leaves us with a system of two linear equations. Solving gives Eq.~(\ref{screened_2d_pot}),
where we note that the Thomas-Fermi screening wavevector $k_{TF}$ 
is defined to be twice what is typical
for GaAs, due to the extra valley degeneracy in Si, and we have assumed that $k \ll k_{TF}$.

\section{Calculation of the power transfer}\label{energy_calculation}
This appendix presents the derivation of Eq.~(\ref{PbyAversion1}).
The calculation begins
similarly to those done in the case of Coulomb drag.\cite{Rojo:1999p10} 
However, it proceeds quite differently because the symmetry of the momentum transfer relevant to CD differs from that of power transfer, which we consider here.
Following the CD literature,\cite{Rojo:1999p10} we seek to decouple the $k_1$ and $k_2$ integrals. 
First, it is conventional to split the energy-conserving delta function in Fermi's golden rule by introducing an integration over the transfer energy. The relevant identity is
\cite{Rojo:1999p10}
\begin{align}
& \delta \left( E_{\mathbf k_1} +E_{\mathbf k_2} - E_{\mathbf k_1+\mathbf k}-E_{\mathbf k_2-\mathbf k} \right)   \\
 &= \int d E \delta(E +  E_{\mathbf k_1} - E_{\mathbf k_1+\mathbf k}) \delta(E-E_{\mathbf k_2} +E_{\mathbf k_2-\mathbf k} ). \nonumber
\end{align}
Next, note that Fermi-Dirac distributions at a common temperature $T$ satisfy the algebraic relationship for two energies $E_x$ and $E_y$ \cite{Rojo:1999p10}
\begin{equation}
f(E_x) \left( 1-f(E_x+E_y) \right) = \frac{f(E_x)-f(E_x+E_y)}{1-e^{-E_y/(k_B T)}}.
\end{equation}
A third, useful algebraic identity is
\begin{equation}
\frac{1}{1-e^{-a}}\frac{1}{1-e^{+b}} -\frac{1}{1-e^{+a}}\frac{1}{1-e^{-b}}  = \coth b - \coth a,
\end{equation}
which is verified by using the definition of the hyperbolic tangent. With these identities, it is tedious but straightforward to show that Eq.~(\ref{eq:formal_energy_transfer}) can be written as
\begin{align} \label{power_intermediate}
\frac{P}{A} & =  \frac{16 q^2}{\hbar (2 \pi)^5} \int d ^2 k dE \cdot 
 E  \left|\tilde \phi_{tot}(k,d) \right|^2  I(\mathbf k, E) J(\mathbf k, E) \nonumber \\
&\times  \left[ \coth \left( \frac{E}{k_B T_2} \right) -\coth \left( \frac{E}{k_B T_1} \right) \right],
\end{align}
where
\begin{align}
I(\mathbf k, E) & = \int d^2 k_1  \delta(E +  E_{\mathbf k_1} - E_{\mathbf k_1+\mathbf k})  \nonumber \\
& \times \left[ f^{(1)}(E_{\mathbf k_1}) - f^{(1)}(E_{\mathbf k_1}+E)  \right] 
\end{align}
and 
\begin{align}
J(\mathbf k, E) & =\int d ^2 k_2 \delta(E-E_{\mathbf k_2} +E_{\mathbf k_2-\mathbf k} ) \nonumber \\
& \times \left[ f^{(2)}(E_{\mathbf k_2}) - f^{(2)}(E_{\mathbf k_2}-E) \right].
 \end{align}
 Again, $f^{(1)}$ is the Fermi-Dirac distribution function of layer 1 and $f ^{(2)}$ is the Fermi-Dirac distribution function
 of layer 2.
 We next make the simplifying assumption that the carriers in the two layers have have the same effective masses, but possibly different Fermi levels.  Assuming that the temperature is sufficiently low, we also approximate the distribution functions as step functions at the
 Fermi level, from which it follows that $I(\mathbf k,E) \approx - J(\mathbf k, E)$ when the Fermi levels are the identical. 

 We calculate $I(\mathbf k, E)$ within the effective mass approximation with a simple parabolic dispersion, $E = \hbar^2 k^2/(2 m^*)$, by using Cartesian coordinates and integrating over $k_1$, yielding:
\begin{equation}
I(\mathbf k, E) =  \frac{m^*}{\hbar^2 k}\sqrt{\frac{m^*}{2 \hbar^2}} \mathrm{Re} \left( \sqrt{E_\beta} - \sqrt{E_\alpha} \right),
\end{equation}
where $m^*$ is the (transverse) effective mass, $E _\alpha = E_F-E_0-E$, $E _\beta = E_F-E_0$, $E_F$ is the Fermi level of layer 1 and
\begin{equation}
E_0 = \frac{\hbar^2}{2m^*} \left( \frac{k}{2}-\frac{E m^*}{\hbar^2 k}\right)^2.
\end{equation}
It is now useful to switch to dimensionless coordinates, where we define $\zeta \equiv E/E_F$ and $\eta \equiv k/k_F$, where $k_F = \sqrt{2 m^* E_F}/\hbar$ is the Fermi momentum. Doing this gives
\begin{align}\label{I_integral}
I = \frac{m^*}{4 \hbar^2} \cdot \frac{1}{ \eta^2} \mathrm{Re} \Big( &\sqrt{ 2(2 + \zeta) \eta^2 - \eta^4 - \zeta^2} \nonumber \\
 & -\sqrt{ 2(2 - \zeta) \eta^2 - \eta^4 - \zeta^2} \Big).
\end{align}
The calculation of $J$ is very similar, except that the Fermi level of layer two is taken to be $ E_F/x$, where $x = n_1/n_2$ is the ratio of carrier densities.
Recall that we wish to consider systems where $T_1\gg T_2$. Hence, for simplicity we let $T_2 \approx 0$. 
Introducing the parameters $\zeta_0 \equiv k_B T_1/E_F$ and $\eta_0 \equiv 1/(k_F d)$ and substituting
Eqs.~(\ref{screened_2d_pot}) and (\ref{I_integral}) into Eq.~(\ref{power_intermediate}) gives us
\begin{equation}
\frac{P}{A} = \frac{E_F^4}{64 \hbar} \left( \frac{\epsilon_0 \epsilon_b}{q^2} \right)^2 \int d \eta  \left( \frac{\eta}{\sinh(\eta / \eta_0)} \right)^2 Y(\eta, \zeta_0),
\end{equation}
which is Eq.~(\ref{PbyAversion1}), where $Y$ is defined by Eq.~(\ref{yfunc2}). 

\section{Asymptotic analysis of heat transfer}\label{asymptotic}
In this appendix, we seek to obtain an accurate analytic expression for Eq.~(\ref{PbyAversion1}) for both large ($d \gg  E_F/(k_F k_B T)$) and small ($d \ll  E_F/(k_F k_B T)$) separations between the 2DEG layers. 
In the following, we set the densities of the two 2DEGs to be equal for simplicity. 
To, proceed, we first expand for small momentum excitations about the Fermi level. Using this, we work out the asymptotic form of $Y$ (Eq.~(\ref{yfunc2})) on either side of its peak. Then, we calculate the resulting integral in Eq.~(\ref{PbyAversion1}), and derive formulas for asymptotic power transfer for both large (Eq.~(\ref{asymptotic_large_d}))  and small (Eq.~(\ref{asymptotic_small_d})) separations.

\subsection{Asymptotic forms of $Y$}\
To begin, we calculate the asymptotic forms of $Y$.
Since we are at low temperatures, we may assume that the transfer momentum $k \ll k_F$, the Fermi momentum. In this approximation, we find that
\begin{align}\label{yapprox}
Y(\zeta,\eta_0) & \approx \int_0^{2 \eta - \eta^2} \frac{\zeta(\coth(\zeta/\zeta_0)-1)}{\eta^3}\frac{4 \zeta^2 \eta^2}{4-\zeta^2/\eta^2}  \\
& + \int_{2 \eta - \eta^2}^{2 \eta + \eta^2} \frac{\zeta(\coth(\zeta/\zeta_0)-1)}{\eta^3}\left[ 2(2+ \zeta) \eta^2-\eta^4-\zeta^2 \right]. \nonumber
\end{align}

The two limiting cases we consider are when $\eta \ll \zeta_0$, corresponding to the region well to the left of the peak in $Y$, and $\eta \gg \zeta_0$, corresponding to the right of the peak. When  $\eta \ll \zeta_0$, we make the approximation that
\begin{equation}
\coth(\zeta/\zeta_0) \approx \zeta_0 - \zeta.
\end{equation}
Using this, Eq.~(\ref{yapprox}) reduces to
\begin{equation}
Y(\eta,\zeta_0) \underset{\eta \ll \zeta_0}{\sim} 4 \zeta_0 \log (4/\eta)\eta^2.
\end{equation}
When  $\eta \gg \zeta_0$, we may take $\eta \gg \zeta$, so
\begin{equation}
\frac{3 \zeta^2 \eta^2}{4-\zeta^2/\eta^2} \approx \zeta^2 \eta^2.
\end{equation}
In this limit, the second integral in Eq.~(\ref{yapprox}) does not contribute. Evaluating the first integral, we find
\begin{equation}
Y(\eta,\zeta_0) \underset{\eta \gg \zeta_0}{\sim} \frac{\zeta_0^4}{\eta} \int_0^\infty dx (\coth x - 1 )x^3,
\end{equation}
which reduces to
\begin{equation}\label{yasymptote2}
Y(\eta,\zeta_0)  \underset{\eta \gg \zeta_0}{\sim} \frac{\zeta_0^4}{\eta}\frac{\pi^4}{120}.
\end{equation}
Now that we have the asymptotic forms of $Y$ on either side of the peak, we may proceed to evaluate Eq.~(\ref{PbyAversion1}) in the limits of
$\eta_0 \ll \zeta_0$ and $\eta_0 \gg \zeta_0$. 

\begin{figure}[tb!] 
\begin{center} 
\includegraphics[width=0.9 \linewidth]{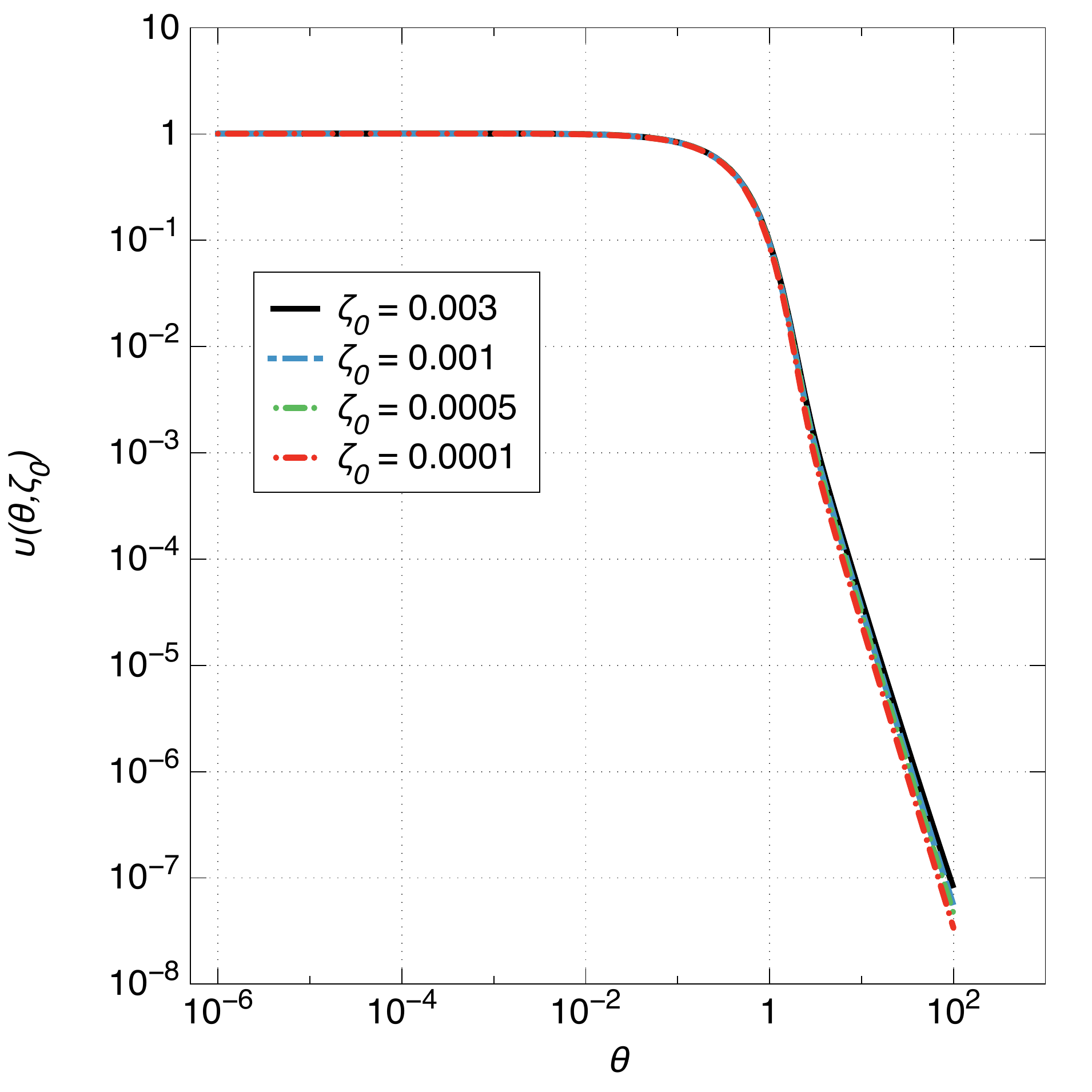}
\end{center} 
\caption{(Color online) The rescaled function $\upsilon(\theta)$ (Eq.~(\ref{upsilondef})), plotted versus the scaled coordinate $\theta = \eta/\zeta_0$. Here, $\eta = k/k_F$, the momentum transfer scaled by the Fermi momentum, and $\zeta_0 = k_B T / E_F$. For small $\theta$, $\upsilon$ takes
on the expected asymptotic value of one. The curves essentially coincide until around $\theta \gtrsim 1$. This enables us to treat $\upsilon$ as 
approximately independent of $\zeta_0$ before that point.
\label{fig_5} }
\end{figure}

\subsection{Power transfer in the limit of large separation}
Next, we evaluate Eq.~(\ref{PbyAversion1}) in the large-distance limit, when $\eta_0 \ll \zeta_0$, which corresponds to  $d \gg  E_F/(k_F k_B T)$.
To do this, we first define a scaled function $\upsilon$:
\begin{equation}\label{upsilondef}
\upsilon(\theta,\zeta_0) \equiv \frac{Y(\theta \zeta_0,\zeta_0)}{4 \zeta_0^3 \log(4/(\theta \zeta_0)) \theta^2},
\end{equation}
which is just $Y$ scaled by its asymptotic value in the region where $\eta \ll \zeta_0$ as a function of the scaled coordinate $\theta \equiv \eta/\zeta_0$. We plot $\upsilon$ for various values of $\zeta_0$ in Fig.~\ref{fig_5}.

The integral we need to evaluate can be written as
\begin{equation} \label{Lambdadef}
\Lambda(\eta_0,\zeta_0) \equiv 4 \zeta_0^6  \int_0^\infty d\theta \frac{\theta^4\log(4/(\theta \zeta_0))\upsilon(\theta,\zeta_0)}{\sinh^2(\theta \zeta_0/\eta_0)},
\end{equation}
which is related to Eq. \ref{PbyAversion1} by
\begin{equation} \label{PbyAversion2}
\frac{P}{A} = \frac{E_F^4}{64 \hbar} \left( \frac{\epsilon_0 \epsilon_b}{q^2} \right)^2 \Lambda(\eta_0,\zeta_0).
\end{equation}
Since we are in the region where $\eta_0 \ll \zeta_0$, we can approximate $\upsilon(\theta,\zeta_0) \sim 1$. The integration
can then be carried out numerically, resulting in
\begin{equation}
\Lambda(\eta_0,\zeta_0)  \underset{\eta_0 \ll \zeta_0}{\sim} \zeta_0 \eta_0^5 \left( 8.3 - 13.0 \log \eta_0 \right),
\end{equation}
which reduces to Eq.~(\ref{asymptotic_large_d}) when inserted into Eq.~(\ref{PbyAversion2}).

\subsection{Power transfer in the limit of small separation}
Now, we evaluate Eq.~(\ref{PbyAversion1}) in the small-distance limit, when $\eta_0 \gg \zeta_0$, corresponding to  $d \ll  E_F/(k_F k_B T)$.
We begin with Eq.~(\ref{Lambdadef}), but unlike before we cannot assume that $\upsilon(\theta, \zeta_0) = 1$. Instead, we note that from Fig.~\ref{fig_5}, $\upsilon$ is approximately independent of $\zeta_0$ until some cutoff $\theta$, $\theta_c \gtrsim 1$. Hence, we split the
integration region of $\Lambda$ into two pieces at $\theta_c$. For $\theta < \theta_c$, we take $\upsilon$ to be independent of $\zeta_0$, and 
also
\begin{equation}
\sinh(\theta \zeta_0/\eta_0) \approx \frac{\theta \zeta_0}{\eta_0},
\end{equation}
where this second approximation is valid since $\theta \leq \theta_c \ll \eta_0/\zeta_0$. For $\theta>\theta_c$, we calculate $\upsilon(\theta,\zeta_0)$ according to the asymptotic formula for $Y$ in the limit of $\eta \gg \zeta_0$, given in Eq.~(\ref{yasymptote2}). These approximations result in
\begin{align}
\Lambda(\eta_0,\zeta_0) & \underset{\eta_0 \gg \zeta_0}{\sim} 
4 \zeta_0^4 \eta_0^2 \int_0^{\theta_c} d\theta \log(4/(\theta \zeta_0))\theta^2\upsilon(\theta) \nonumber \\
&+ \frac{\pi^4}{120}\zeta_0^6 \int_{\theta_c}^\infty d \theta \frac{\theta}{\sinh^2(\theta \zeta_0/\eta_0)}.
\end{align}
In the limit where $\zeta_0 \ll \eta_0$ and with $\theta_c = 2.0$, the result is
\begin{equation}
\Lambda(\eta_0,\zeta_0)   \underset{\eta_0 \gg \zeta_0}{\sim} \zeta_0^4 \eta_0^2 \left( 0.458 - 1.32 \log \zeta_0 + 0.182 \log \eta_0 \right).
\end{equation}
In this evaluation, we picked $\zeta_0 = 0.001$ for the calculation of $\upsilon$ in the region where it is approximately $\zeta_0$-independent.
Inserting this result into Eq.~(\ref{PbyAversion2}) gives Eq.~(\ref{asymptotic_small_d}), as desired.

\bibliography{refs}
\end{document}